**Title:** A robust approach for spectral analysis: "optimized spectral replacement"

**Applicant:** Dr Mansour Taghavi Azar Sharabiani

**Layman's summary**

Metabonomics [1, 2], the measure of the fingerprint of biochemical perturbations caused by disease, drugs or toxins, recently has become a major focus of research in various areas [3] especially indications of drug toxicity [4-8]. Two types of technology (known by the initials NMR and MS) are employed and both produce massive data in form of spectra. Sophisticated statistical models, known as pattern recognition techniques, are commonly applied for summarizing and analysing these multidimensional data. However, strong signals from compounds that are administered during toxicological trials interfere with these models. So called "spectral replacement" [9] is a method to eliminate these signals by replacing them with the signals in their corresponding regions in control spectrum. The replaced regions are subsequently scaled. However, this scaling is not accurately measured and often results in overestimation of integrated intensity of the replaced signals. Here, a novel protocol is proposed which provides an accurate estimation of the replaced regions.

**Background**

Cells and tissues attempt to recover from 'disease- or xenobiotic-induced' disturbances of ratios, concentrations, and fluxes of chemicals and molecular species (metabolites) via dynamically modulating the biochemical composition of the body fluids through perfusion or secretion. Therefore, biofluids or tissue samples potentially contain a wealth of information regarding metabolic status (concentrations of various metabolites) of organisms and may be indicative of toxicity or disease conditions. For this reasons, metabonomics, defined as "quantitative measurement of time-related multi-parametric metabolic responses of multi-cellular systems to pathophysiological stimuli or genetic modification" [1, 2], recently has become a major focus of research in various areas [3] especially indications of drug toxicity [4-8]. Xenobiotics are compounds such as toxins or drugs that are not normally present in organisms or are at much lower levels before administration. Metabolites are small molecules, normally present in all organisms, and are essential to the functioning of their living cells. Metabonomics originates from metabolite profiling [10], which initially appeared in literature in 1950s [7]. Metabonomics employs two specific types of technology i.e. NMR and MS, which are sophisticated procedures to measure concentrations of compounds in samples and both produce a large amount of data in form of spectra. Each spectrum contains massive information about the profiles of metabolites and xenobiotics (if present). In Figure 1 there are five different NMR spectra of urine samples taken from five rats, which four of them received different treatments and one of them (at the bottom) did not (control). Red arrows indicate strong signals from administered xenobiotics (drugs).

For the rest of the text, spectra of samples from organisms subjected to xenobiotic treatments are referred to as 'test spectra' and those without treatment as 'control spectra'. Also, those regions of test spectrum in which the drug metabolite signals appear are referred to as $C$ regions (see Figure 2(6)) and their corresponding regions in control spectra as $c$. Similarly, regions of indigenous metabolite signals in test spectrum referred to as $B$ (see Figure 2(3)) and their corresponding regions in control as $b$. The rest of test spectrum (without $C$ and $B$) is referred to as $A$ (see Figure 2(2)) and their corresponding regions in control as $a$. Figure 2 illustrates two simple computationally simulated spectra (test and control shown by red and black lines, respectively). Two simulated spectra are superimposed and therefore the control spectrum (black line) is visible only in those regions wherein two spectra do not overlap i.e. $B$ (or $b$) regions.

Expert systems capable of predicting the likelihood of the toxicity of novel drug candidates have been developed employing series of mathematical models. Pattern recognition techniques such as principle component analysis (PCA) [11] are widely used for summarising these multidimensional data. PCA captures and displays the most dominant patterns in the data matrix. However, xenobiotic compounds administered during toxicological trials to the organisms often generate strong signals, which interfere with these mathematical models, presenting themselves as the most dominant patterns.

**Figure 1 Five NMR spectra of urine samples from rats which four of them received different treatments and the last one (at the bottom) did not receive any treatment. Strong signals from administered xenobiotics are indicated with red arrows**

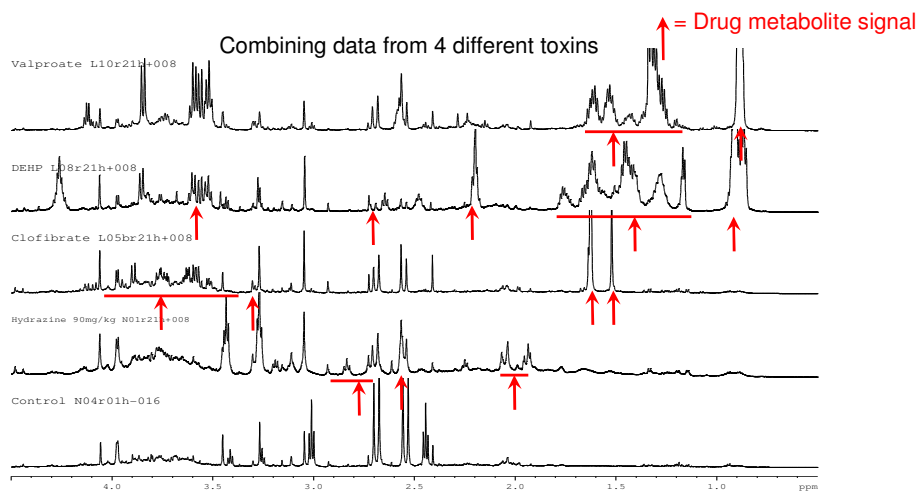

**Figure 2 Two simple simulated test and control spectra**

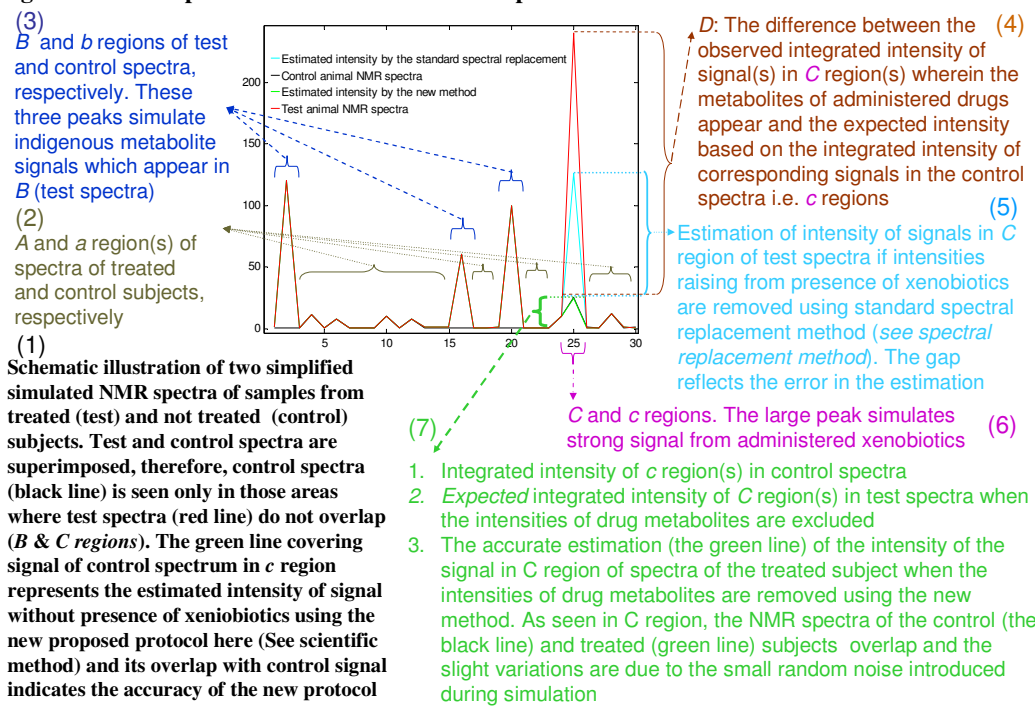

(3) *B* and *b* regions of test and control spectra, respectively. These three peaks simulate indigenous metabolite signals which appear in *B* (test spectra)

(2) *A* and *a* region(s) of spectra of treated and control subjects, respectively

(1) **Schematic illustration of two simplified simulated NMR spectra of samples from treated (test) and not treated (control) subjects. Test and control spectra are superimposed, therefore, control spectra (black line) is seen only in those areas where test spectra (red line) do not overlap (*B* & *C* regions). The green line covering signal of control spectrum in *c* region represents the estimated intensity of signal without presence of xeniobiotics using the new proposed protocol here (See scientific method) and its overlap with control signal indicates the accuracy of the new protocol**

(4) *D*: The difference between the observed integrated intensity of signal(s) in *C* region(s) wherein the metabolites of administered drugs appear and the expected intensity based on the integrated intensity of corresponding signals in the control spectra i.e. *c* regions

(5) Estimation of intensity of signals in *C* region of test spectra if intensities raising from presence of xenobiotics are removed using standard spectral replacement method (*see spectral replacement method*). The gap reflects the error in the estimation

(6) *C* and *c* regions. The large peak simulates strong signal from administered xenobiotics

(7)
1. Integrated intensity of *c* region(s) in control spectra
2. *Expected* integrated intensity of *C* region(s) in test spectra when the intensities of drug metabolites are excluded
3. The accurate estimation (the green line) of the intensity of the signal in C region of spectra of the treated subject when the intensities of drug metabolites are removed using the new method. As seen in C region, the NMR spectra of the control (the black line) and treated (green line) subjects overlap and the slight variations are due to the small random noise introduced during simulation

Since most of xenobiotic metabolites and their spectral regions (i.e. *C* regions) are well known, it would have been possible to simply excise all these regions if there were a small number of toxins in a database with a scarce number of xenobiotic-related peaks. When one region is excised, all corresponding spectral regions of the excised part in all other spectra must also be excised so as to keep them comparable. This can pose a major problem for large databases such as the COMET[12, 13] project containing wide range of model toxins. In these situations, almost all of the spectral regions might have to be eliminated (See Figure 1). Other procedures like replacing these regions with zero or mean values are not optimum alternative when there are large numbers of spectra. Thus, it is necessary to somehow estimate the 'increased integrated intensity' of signals in *C* regions, which are 'added' purely due to presence of administered xenobiotics (See Figure 2 (4)). So called "spectral replacement" [9] is a method which has been developed for this purpose and applied in a number of studies [14-16].

**Scientific methodology**

As mentioned earlier, strong signals of the administered xenobiotics obscure signatures of indigenous metabolic changes which are characteristics of organisms' responses to the xenobiotics. Spectral replacement is a bioinformatics solution to eliminate xenobiotic signals by replacing them with their corresponding regions in control spectrum. In other words, $C$ regions of test spectrum is replaced with corresponding $c$ region from control spectrum. The inserted $c$ region in test spectrum is scaled by a factor, $f$, using Equation 1 so as to have the same fraction of the total integrated intensity of test spectrum as it did in said control spectrum Figure 2 (5). Since in this method, the fraction of integrated intensity of replaced $c$ regions to total integral intensity of the test spectrum is estimated based on their fraction to total integral intensity of control spectrum, the influence of the significant changes of indigenous metabolites (as result of toxic reaction) on total integrated intensity of test spectrum is not taken into account. This leads to inaccurate estimation (most likely overestimation) of integrated intensity of replaced signals, see Figure 2 (4, 5). The novel protocol proposed here provides an accurate estimation of integrated intensity of replaced signals in test spectrum; see Figure 2(1, 7). Figure 2 schematically illustrates two superimposed test and control spectra, red and black lines, respectively. Turquoise and green lines represent estimated integrated intensities of the replaced signals in $C$ region using spectral replacement and the novel method, respectively.

**Equation 1 extracted from spectral replacement[9]**

$$f = \frac{I_Y - \sum_k I_{Y,Tk}}{I_{CM} - \sum_k I_{CM,Rk}}$$

wherein:
$I_Y$ is the total integrated intensity of the test spectrum;
$I_{Y,Tk}$ is the integrated intensity of the target (i.e. $C$) region;
$I_{CM}$ is the total integrated intensity of the control spectrum;
$I_{CM,Rk}$ is the integrated intensity of the replacement region i.e. $c$ region;
k ranges from 1 to nt; and,
nt is number of target (i.e. $C$) regions.

To simplify comparison, in this simulation it is assumed that the test and the control spectra are from two samples with identical molar concentrations, therefore, ideally the estimated integrated intensity of signals in $C$ region of the test spectrum must have the same integrated intensity of $c$ region of the control spectrum, i.e. the green and the black lines should overlap (see Figure 2 (1, 6, 7)). According to biological assumptions, $A = ka$ and $B = Kb + I$, where $I$ is the amount of changes of integrated intensity of $B$ regions after xenobiotics administration and $k$ is a scaling factor reflecting difference between concentrations of the biofluid samples. Similarly, $C = Kc + D$, where $D$ is the 'added' intensity of signals in $C$ region resulting from the presence of administered xenobiotics. Therefore, test ($Test$) and control ($Ctrl$) spectra can be defined as:

$Ctrl = a + b + c$ and $Test = ka + kb + I + kc + D$

If the replacement and scaling measure is carried out correctly, the ideal spectrum, $Goal$, should be a specific version of test spectrum in which $D$ (see Figure 2(4)), has been subtracted from integrated intensity of the $C$ regions, while $I$ integrated intensity retained in $B$ regions. Therefore:

$Goal = ka + kb + kc + I$ or $Goal = k(a+b+c+i)$

The $Goal$ spectrum can be calculated via the following proposed four-step protocol:

*STEP I*: $C$ and $c$ regions are removed from test and control spectra respectively.

$Ctrl_{STEPI} = a + b$ and $Test_{STEPI} = ka + kb + ki$

*STEP II*: $b$ regions of the step 1 control spectrum are replaced with the corresponding $B$ regions of the step 1 test spectrum. Each replaced region is scaled to a factor of $F_1$ using Equation 1 so as to have the same fraction to the total integrated intensity of the step 1 control spectrum (i.e. after removal of $c$ regions) as it did it in the step 1 test spectrum (i.e. after removal of $C$ regions).

Thus $Ctrl_{STEPII} = a + F_1(k(b+i))$

According to Equation 1, we know $\dfrac{F_1(k(b+i))}{a + F_1(k(b+i))} = \dfrac{k(b+i)}{k(a+b+i)}$

The rearrangement of this equation gives $F_1 = \dfrac{1}{K} \Rightarrow Ctrl_{STEPII} = a + b + i$

*STEP III*: Deleted *c* regions are returned to control spectrum $Ctrl_{STEPIII} = a + b + i + c$

*STEP IV*: Deleted *c* regions of the control spectrum are inserted into the sample spectrum in place of deleted corresponding *C* regions. Inserted *c* regions are scaled to a factor of $F_2$ using Equation 1 so as to have same fraction of total integrated intensity of test spectrum as they did in step 3 control spectrum
$Test_{STEPIV} = k(a + b + i) + F_2 c$

According to Equation 1, we know $\dfrac{F_2 c}{k(a+b+i+c) + F_2 c} = \dfrac{c}{a+b+i+c} \Rightarrow F_2 = k$

Thus: $Test_{STEPIV} = k(a + b + i + c) = Goal$

*The main objectives of this project are:*
- Development of bioinformatics tool for implementation of the novel protocol for properly eliminating strong signals of xenobiotics from spectrum
- Comparing the results from both simulated and biological data to estimate the performance of the software and the protocol in 'real world' applications
    - A procedure will be developed to compare the effect of implementation of the new protocol, spectral replacement, and other techniques such as zero-filling and mean value-filling on the results of pattern recognition models and techniques such as PCA using both simulated and biological data

**Timeliness and novelty**

Research and development cost of new drugs continues to rise above general price inflation while approval rates fall [17]. Efficient and reliable experimental / analytical chemistry approaches to eliminate xenobiotic metabolites are yet to be developed. Application of experimental methodologies such as labelling (fluorescent or radioactive) and NMR-HPLC techniques for this purpose can be time consuming and expensive with excessive bio- and radioactive hazards to environment as well as increased use of animals in experiments which raises ethical concerns. Fluorescence labelling may interact with xenobiotics. In those circumstances where xenobiotics normally exits in organism at lower levels before administration, even combination of radiolabelling and HPLC-NMR techniques also cannot separate co-eluting radiolabelled/organism's xenobiotics. By contrast, bioinformatics approach proposed here, backed with mathematical proof, provides a cheap, rational, and relatively quick solution to eliminate the problem of strong xenobiotic signals. This protocol could be applied to other similar areas of data analysis. This project could also be viewed as is a measure of rectifying the inaccuracy of spectral replacement method, which has already been applied in a number of major metabonomic studies. Therefore this protocol could potentially play a major role in analysis and modelling of NMR and other forms of spectral data. The product of this project and the novel protocol itself could be potentially patentable.

**Programme of work**

This is a short term project and the following tasks are expected to be accomplished during 6 months period (Simulations will be performed within Matlab environment, Data from COMET project [12] as well as publicly available databases will be used):

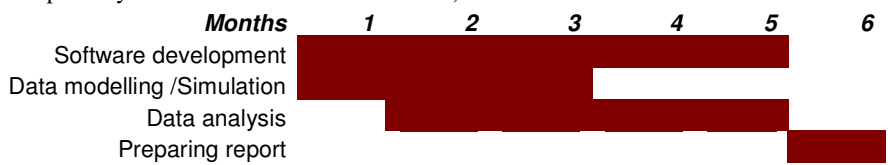

**Justification of resources**

- Principle investigator: 6 months graduate student / research assistant
- Desktop PC (a platform for developing software and data analysis, and preparing final report)
- Software Licenses, including Matlab, SIMCA-P and NMR processing software
- Consumables - Stationery, photocopying charges, toner library charges, etc

**Beneficiaries**

Aiming to resolve one of the major confounding factors in modelling of data from metabonomic spectra, this small project potentially has a wide range of beneficiaries including
- The growing metabonomic community who analyse NMR and other types of spectra
- Scientists involved in developing or using databases of NMR or other types of spectra
- With expansion of metabonomic applications such as pharmaco-metabonomics, nutritional-metabonomics system biology, involved scientists such as system biology scientists will have more accurate information, as the result of this project, to integrate in holistic modelling of fundamental biological processes
- Pharmaceuticals, food and agriculture industries, those involved in toxicological studies and those for which understanding and monitoring metabolic activity and quality control of drugs or food is essential, Researchers involved in preclinical and clinical toxicological screening as well as those involved in designing personalised treatments, or diets

Because the project increases the better use of existing information and also improves the use of metabolic profiling in screening procedures, it will indirectly contribute to the reduced use of animals in biological research and also reduced bio and radioactive hazards to environment